\documentclass[conference,onecolumn,10pt,letterpaper]{IEEEtran}
\IEEEoverridecommandlockouts
\usepackage{cite}
\usepackage{amsmath,amssymb,amsfonts}
\usepackage{algorithmic}
\usepackage{graphicx}
\usepackage{textcomp}
\usepackage{xcolor}
\def\BibTeX{{\rm B\kern-.05em{\sc i\kern-.025em b}\kern-.08em
    T\kern-.1667em\lower.7ex\hbox{E}\kern-.125emX}}

\usepackage{listings}

\usepackage{pgfplots}
\usepackage{pgfplotstable}
\usepackage{amssymb}
\usepackage{comment}
\usepackage{xcolor}
\usepackage{etoolbox}
\newbool{inccomment}
\setbool{inccomment}{true}
\definecolor{darkgreen}{RGB}{0,140,0} %

\newcommand{\sh}[1]{\ifbool{inccomment}{{\color{blue}#1}}{}}
\newcommand{\bo}[1]{\ifbool{inccomment}{{\color{darkgreen}#1}}{}}

\definecolor{vgreen}{RGB}{104,180,104}
\definecolor{vblue}{RGB}{49,49,255}
\definecolor{vorange}{RGB}{255,143,102}

\lstdefinestyle{verilog-style}
{
	language=Verilog,
	basicstyle=\small\ttfamily,
	keywordstyle=\color{vblue},
	identifierstyle=\color{black},
	commentstyle=\color{vgreen},
	numbers=left,
	numberstyle=\tiny\color{black},
	numbersep=10pt,
	tabsize=8,
	moredelim=*[s][\colorIndex]{[}{]},
	literate=*{:}{:}1
}
\usepackage{pgfplots}
\usepackage{caption}
\usepackage{subcaption}
\usepackage{geometry}
\usepackage{comment}

\definecolor{burgundy}{rgb}{0.5, 0.0, 0.13}
\definecolor{blush}{rgb}{0.87, 0.36, 0.51}
\definecolor{ballblue}{rgb}{0.13, 0.67, 0.8}
\definecolor{TUMBlau}{RGB}{0,101,189} 
\definecolor{TUMBlauDunkel}{RGB}{0,82,147} 
\definecolor{TUMBlauHell}{RGB}{152,198,234} 
\definecolor{TUMBlauMittel}{RGB}{100,160,200} 
\definecolor{TUMElfenbein}{RGB}{218,215,203} 
\definecolor{TUMGruen}{RGB}{162,173,0} 
\definecolor{TUMOrange}{RGB}{227,114,34} 
\definecolor{TUMGrau}{gray}{0.6} 
\usepackage{caption}
\usepackage{ulem}
\begin{document}

\title{Automated Formal Verification of Area-Optimized Safety Registers in Automotive SoCs}

\author{
Shuhang Zhang\IEEEauthorrefmark{1},
Bryan Olmos\IEEEauthorrefmark{1}\IEEEauthorrefmark{2},

\\


\IEEEauthorrefmark{1}Infineon Technologies AG\\
\IEEEauthorrefmark{2}Rheinland-Pfälzische Technische Universität Kaiserslautern-Landau \\
\emph{Email: Firstname.Lastname@infineon.com}
}

\maketitle

\begin{abstract}
	\textbf{Registers are primary storage elements in System-on-chip~(SoC) designs and play an important role in maintaining state information and processing data in digital systems.  With respect to the ISO26262 standard, these registers require high levels of reliability and fault tolerance. For this reason, safety-critical applications require that normal registers are equipped with additional safety components to construct safety registers, which ensure system stability and fault tolerance. However, the process of integrating these safety registers is complex and error-prone, because of highly-configurable features provided by a safety library such as parameterized modules and flexible safety structures. In addition, to address the overhead caused by the safety registers, we have applied area optimization techniques to their implementation. However, this optimization can make the integration process more susceptible to errors.
    To avoid any integration mistakes, rigorous  verification is always required, but it is time-consuming and error-prone if the  verification is implemented manually when dealing with numerous verification requests. To address these challenges, we propose an automated flow for the  verification of safety registers with the formal approach. 
    The results indicate that this automated verification approach has the potential to reduce the verification effort by more than 80\%. Additionally, it ensures a comprehensive examination of every  requirement of this safety library, which is reflected in faster detection of bugs. The proposed framework can be replicated for the verification of other safety components enabling an early detection of potential  issues and saving valuable time and resources.}
\end{abstract}

\section{Introduction}

In the modern System-on-Chip (SoC) design, registers hold a paramount position, because these fundamental components not only act as the primary storage units within digital systems but also maintain state information. For automotive applications, registers are also required to operate under strict safety and reliability standards, e.g. ISO26262~\cite{ISO26262-6}, which necessitates a high degree of consistency and fault tolerance. To meet these stringent requirements, the ordinary registers are usually armed with safety components to ensure that automotive chips can tackle faults effectively and maintain their operation without compromising the safety and reliability of the overall system. To implement these safety features in registers, a specific safety library, named Safety Flip Flop~(SFF) library~\cite{busch2016automated}, is developed and utilized, providing different safety components to implement various safety features within the register implementation.

However, the integration of these safety registers is also an error-prone process due to the flexibility provided by the SFF library. Firstly, these safety registers themselves are parameterized and highly configurable. For example, this library supports different types of protection methods for registers, from Error Correction Code~(ECC) to Redundancy, and provides optional self-test features. In addition to safety registers, the safety library provides two additional types of safety components: alarms and controllers. Designers have the flexibility to arrange and combine these components to create various safety structures. However, this configurability, stemming from parameterized modules and flexible safety structures, poses significant challenges in verification.
Furthermore, safety designs often lead to substantial area overhead, which is cost-inefficient. To mitigate this, area-optimization techniques are typically employed. Despite their benefits, the implementation of area-optimized registers introduces additional challenges during the integration process. 
Moreover, manually-implemented  verification is time-consuming, especially when dealing with numerous verification requests, because safety registers are utilized in almost all blocks of SoC designs. Therefore, a more efficient approach is needed in our case.

Recognizing these challenges, we propose an automated workflow for the  verification of safety components in register implementations. This approach aims to replace the traditionally manual verification process, significantly reducing the time and effort required while simultaneously eliminating the potential for human error. By introducing automation, we can dramatically increase both the speed and accuracy of the verification process. More specifically, the contributions of this work include:
\begin{itemize}
	\item	We analyze the safety library and develop a verification plan to cover all  requirements of safety components from the SFF library, as well as area-optimized implementations. Additionally, we demonstrate all defined property classes in this work.
	\item	We propose a novel verification flow to generate the integration properties based on safety specifications and defined property classes automatically. These properties can be verified by commercial verification tools naturally.
	\item	Our results show that the manual effort for the safety  verification can be reduced significantly, from around 5Person-Day (5PD) to 1PD for one safety structure. In the last project, our formal framework detected three integration bugs and two additional bugs in area-optimized implementations. In addition, a comparison with a state-of-the-art work~\cite{busch2023integration}, which  focuses on SFF integration verification, is also discussed from different aspects.
\end{itemize}

The rest of this paper is organized as follows. In Section~\ref{sec:background}, we introduce the SFF library and highlight the motivation for the automated formal verification framework. The proposed formal verification framework will be elaborated in Section~\ref{sec:framework}, and the verification results will be presented in Section~\ref{sec:results}. At last, we conclude this work in Section~\ref{sec:conclusion}.

\section{Background and Motivation}\label{sec:background}
In this section, we present some background knowledge regarding the SFF library, including area-optimized approaches and basic concepts related to the proposed framework. We also highlight the challenges associated with the  verification of safety components.


\subsection{Safety Flip Flop (SFF) Library}\label{sec:background:sff}

A detailed diagram of the SFF library is presented in Fig.~\ref{fig:sff_library}. This library provides three main components for constructing safety register implementations in automotive chips: the SFF Wrapper, the Alarm, and the Controller. These components are explained in detail as follows:

\begin{figure}
	\centering
	\includegraphics[scale=0.4]{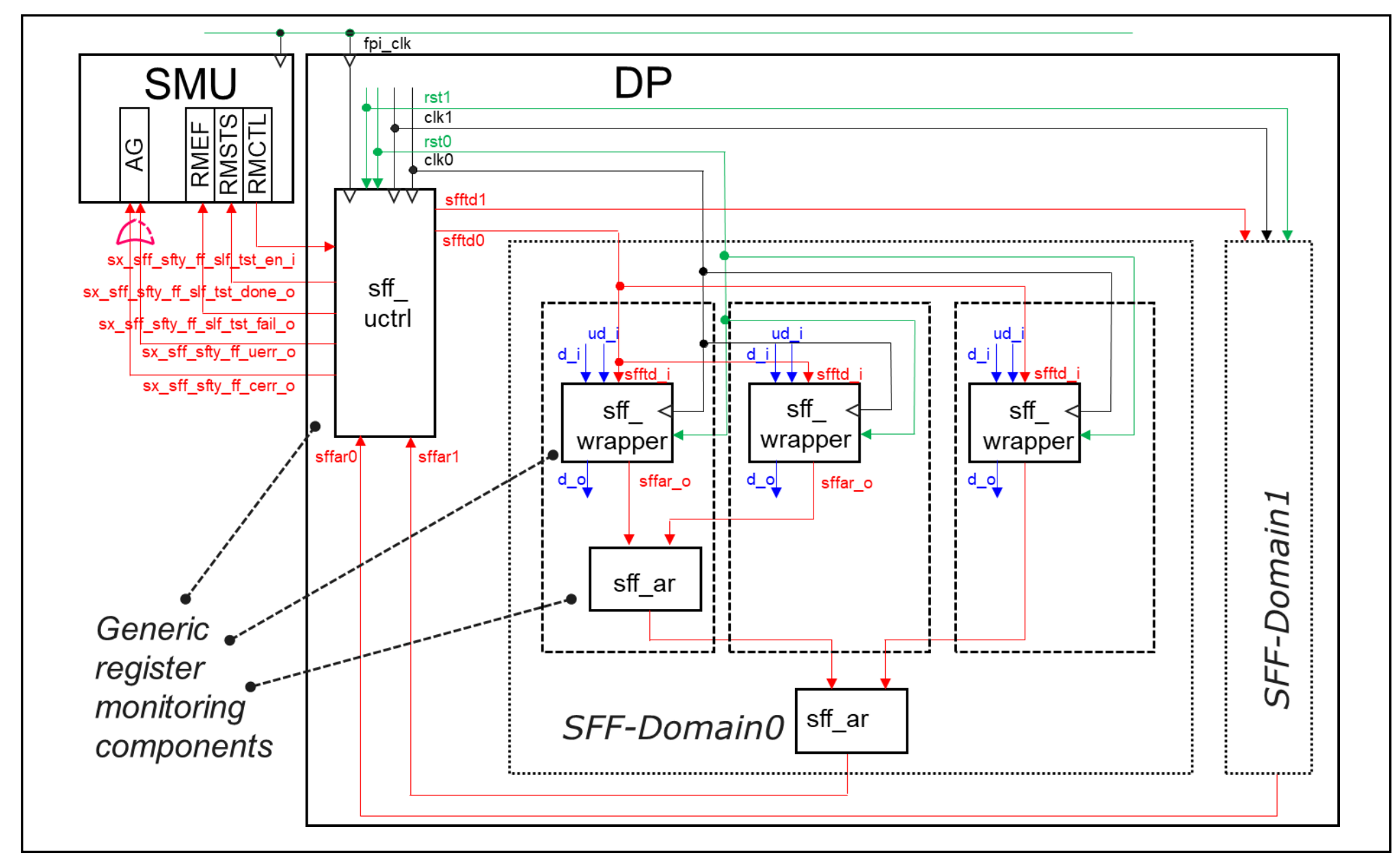}
	\caption{An example of a safety structure with safety components provided by the SFF library.\cite{busch2023integration}}\label{fig:sff_library}
\end{figure}

\begin{itemize}
	\item SFF Wrapper: This is the most important component provided by the library, replacing ordinary register implementations with safety registers. The library supports five different protection methods, classified into two categories: Error Detection and Correction Code (EDCECC)~\cite{hamming1950error}, and Redundancy~\cite{kastensmidt2005optimal}. The specific methods include Parity, Double Error Detection (DED), Single Error Correction Double Error Detection (SECDED), Double Module Redundancy (DMR), and Triple Module Redundancy (TMR). Additionally, optional self-test features~\cite{mccluskey1985built} are supported for the safety registers.
	\item SFF Alarm: This component is used to collect alarm signals from multiple safety wrappers or from different clock domains and generate a reduced alarm signal vector. These reduced alarm signals will be sent to upper-level alarm instances or the SFF Controller.
	\item SFF Controller: This component communicates with the upper-level Safety Management Unit~(SMU). If an error is indicated by the reduced alarm signals, an error flag is sent to the SMU. Additionally, this component controls the self-test features. It receives a trigger signal from the SMU and generates test sequences for all safety registers in different clock domains.
\end{itemize}

\subsection{Area-Optimized Safety Implementation}
\subsubsection{Motivation}
\begin{figure}
\centering
\begin{subfigure}[b]{.45\textwidth}
    \pgfplotsset{compat=1.3,
    /pgfplots/ybar legend/.append style={ 
        /pgfplots/legend image code/.code={%
           \draw[##1,/tikz/.cd,yshift=-0.25em]
           (0cm,0cm) rectangle (5pt,0.8em);
        },
    }
}

\begin{tikzpicture}

\begin{axis}[
xtick={4,64,128,256,512},
ytick={4,5,...,11},
xmin=0, xmax=513,
x=0.012cm, y=0.35cm, 
xticklabel pos=left, xtick align=outside, xtick pos=left,
%
ymin=3.5, ymax=11.5, 
ylabel={$\#$ Redundant Bits}, 
xlabel={Data Width (Bit)},
ylabel style={inner sep=0}, 
%
legend columns=3, 
legend style={
at={(0.9,0.6)}, anchor=south east, font=\tiny,
/tikz/every even column/.append style={column sep=0.2cm},
draw=none, 
},
%
line width=0.75pt,
major tick length=3pt,
]  
    \addplot[sharp plot, line width=1pt, blue!60!black, fill=none] table[x index={0},y index={1},col sep=comma] {fig/sff.csv};

\end{axis}
\end{tikzpicture}
    \vspace{-15pt}
    \caption{Number of redundant bits.}
\end{subfigure}
\hfill
\begin{subfigure}[b]{.45\textwidth}
    \pgfplotsset{compat=1.3,
    /pgfplots/ybar legend/.append style={ 
        /pgfplots/legend image code/.code={%
           \draw[##1,/tikz/.cd,yshift=-0.25em]
           (0cm,0cm) rectangle (5pt,0.8em);
        },
    }
}

\begin{tikzpicture}

\begin{axis}[
xtick={4,64,128,256,512},
ytick={0.4,0.5,...,1},
xmin=0, xmax=513,
x=0.012cm, y=4.7cm, 
xticklabel pos=left, xtick align=outside, xtick pos=left,
%
ymin=0.4, ymax=1, 
ylabel={Code Rate}, 
xlabel={Data Width (Bit)},
ylabel style={inner sep=0}, 
%
legend columns=3, 
legend style={
at={(0.9,0.6)}, anchor=south east, font=\tiny,
/tikz/every even column/.append style={column sep=0.2cm},
draw=none, 
},
%
line width=0.75pt,
major tick length=3pt,
]  
    \addplot[sharp plot, line width=1pt, blue!60!black, fill=none] table[x index={0},y index={2},col sep=comma] {fig/sff.csv};

\end{axis}
\end{tikzpicture}
    \vspace{-15pt}
    \caption{Code rate.}
\end{subfigure}
\caption{The number of redundant bits~(a) and code rate~(b) in a SECDED code implementation with the data width ranging from $4$ to $512$.}\label{fig:secded}
\end{figure}
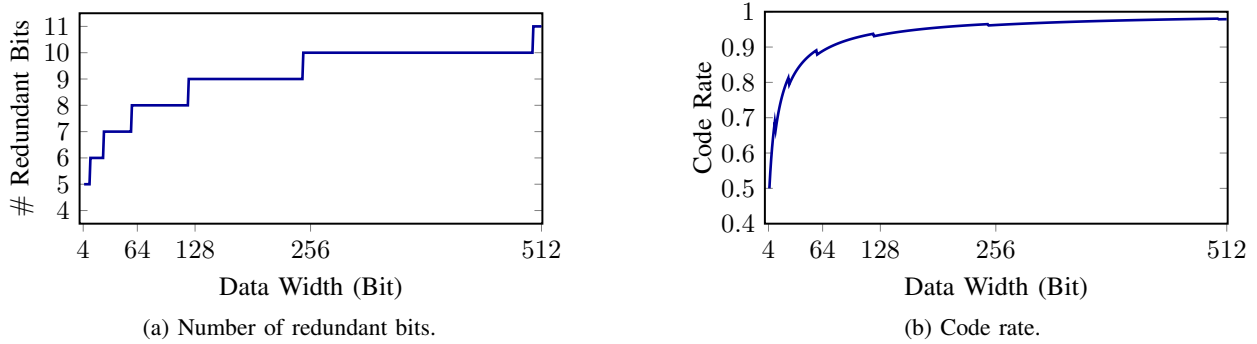

The EDCECC technique is typically employed in safety registers to provide fault-tolerance features, but it also introduces redundant data bits. Given that safety registers consume substantial area, designers usually aim to protect only the critical bits to save area. As a result, a large number of small safety registers are instantiated, which is highly area-inefficient. To illustrate this area inefficiency, we use the SECDED code as an example. The number of redundant bits is shown in Fig.~\ref{fig:secded}(a), and the code rate, defined as $\frac{\# DataBits}{\# TotalBits}$, is shown in Fig.~\ref{fig:secded}(b). These figures clearly indicate that larger safety registers are more area-efficient. To mitigate the area overhead caused by these small registers, merging them into larger ones is a commonly adopted strategy.

\subsubsection{Optimization Strategies}

\begin{figure}

  \includegraphics[width=0.95\textwidth]{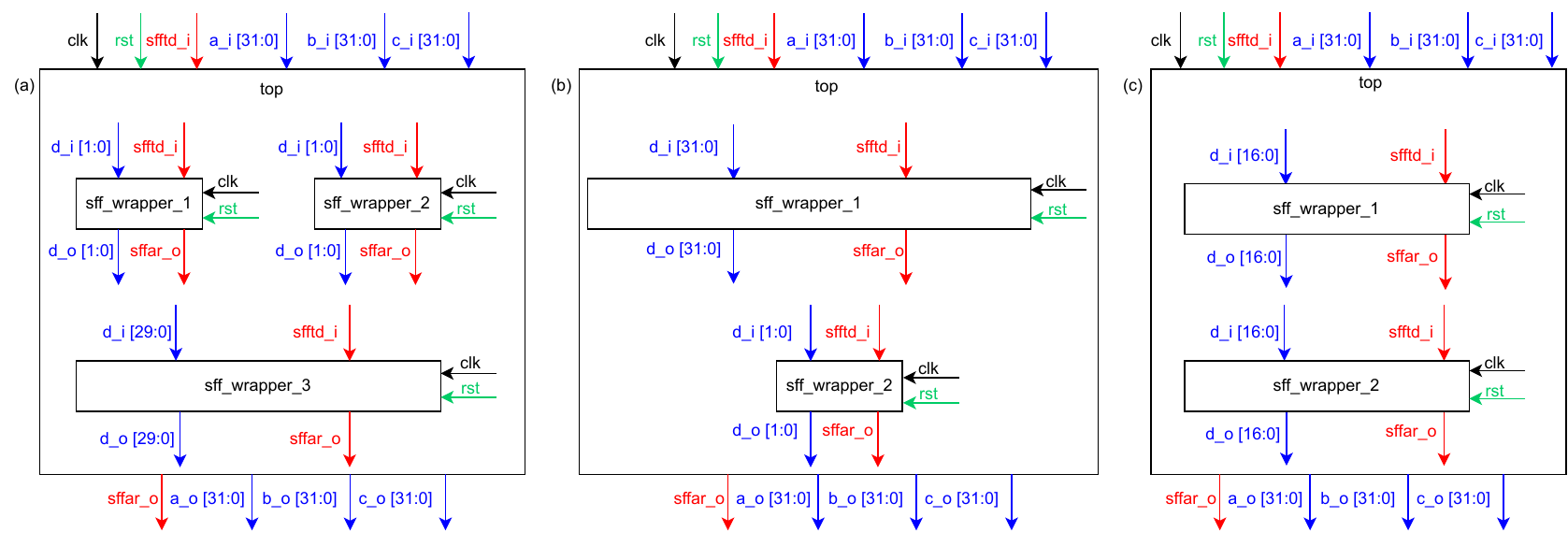}
  \caption{(a) Original design. (b) Optimized design with algorithm 1. (c) Optimized design with algorithm 2.}\label{fig:optimization_data_width}
  \label{fig:example_opt}
\end{figure}

This section explains two optimization approaches based on different algorithms.
The first approach addresses solving a bin-packing problem using the Best-Fit Decreasing~(BFD) technique\cite{kenyon1995best}, which groups small safety registers only considering their size. For example, the original design is shown in Fig.~\ref{fig:example_opt}(a) with three 32-bit registers, labelled as $a$, $b$, and $c$, instantiated within the register block. However, only part of these registers are protected by safety registers: Register $a$ and $b$ have two bits protected each~(sff\_wrapper\_1 and sff\_wrapper\_2), while Register $c$ has 30 bits protected~(sff\_wrapper\_3). We would like to emphasize that in the original design, each register is limited to having only one safety register, and this safety register cannot contain data bits from other registers. Meanwhile, the remaining unprotected bits are not displayed in this figure.  In Fig.~\ref{fig:example_opt}(b), the optimization results of the BFD algorithm are  presented, where the original safety registers are replaced with one large and one small register, considering only the size information of all safety registers. Here, sff\_wrapper\_1, with a data width of 32 bits, includes the data bits from both Register $c$ and $a$. Meanwhile, sff\_wrapper\_2 retains a width of two bits and includes the data bits for Register $b$. 

The second approach uses the Integer Linear Programming~(ILP) technique\cite{schrijver1998theory}, including in the analysis not only the size information of all safety registers but also their synthesized area data.  Fig.~\ref{fig:example_opt}(c) showcases the ILP-optimized design. In this case the algorithm solution is the use of two medium-sized safety registers. In this scenario,  sff\_wrapper\_1 has a data width of 17 bits, protecting part of Register $c$'s data bits. The other safety register contains the remaining bits from Register $c$, along with data bits from Register $a$ and $b$.  
Although the ILP-based approach requires synthesis data during optimization, this is a one-time effort for the same technology, which does not introduce significant additional effort. Consequently, the ILP-based approach can break down large registers and achieve greater area reduction than the BFD-based approach.
Both optimized designs achieve greater area efficiency compared to the original design.

\subsection{Verification Challenges}

As mentioned in the previous sections, the verification challenges are caused by the highly-configurable features. These features can be classified into two parts:
\begin{itemize}
\item 	Parameterized Components and Configurable Safety Structure: Each component from this safety library is configurable, and these components can be combined flexibly to construct various safety architectures to meet different safety requirements. While this flexibility is highly convenient for the design team, it introduces significant challenges for the verification team. To manage the high level of configurability, the formal verification environment must be adapted based on the configurations of the safety architectures being used.
\item   Area-Optimized Implementations: As discussed in the previous section, area-optimized safety registers are necessary to reduce manufacturing costs. Multiple safety registers are merged to form larger ones, thereby reducing area consumption. However, as a result of the merging process, the one-to-one correspondence between signals and registers no longer exists. Additionally, the merging results vary depending on the chosen algorithm, complicating the determination of which signal should connect to the respective parts of the merged registers. Furthermore, the merging is executed automatically by an algorithm, making it essential to verify that the merging adheres to specifications.
\end{itemize}

To address these issues in the safety register verification and achieve exhaustive results, we have proposed an automated formal verification framework.

\section{framework}\label{sec:framework}
\begin{figure}
	\centering
	\includegraphics{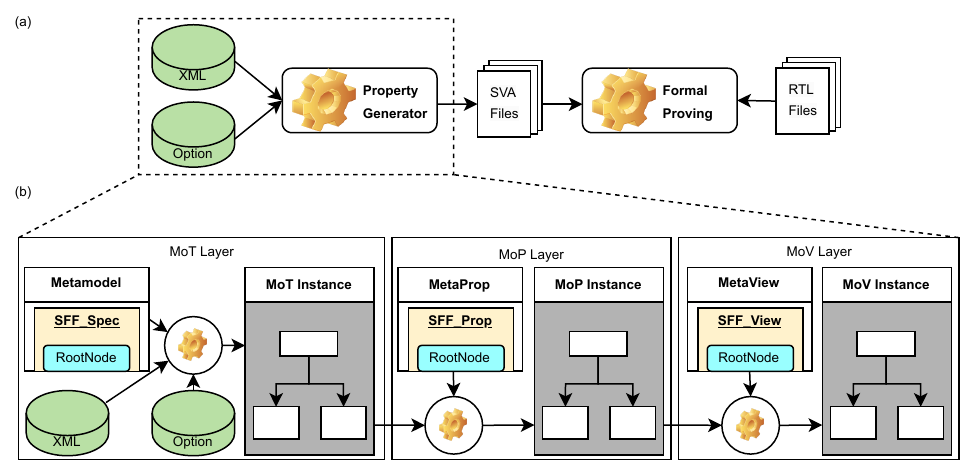}
	\caption{Proposed property generation flow for safety registers.}\label{fig:prop_flow}
\end{figure}

As mentioned in previous sections, the highly-configurable nature of the safety library, along with the area optimization techniques, makes the verification of safety registers a challenging task. Therefore, we primarily address this task with our proposed verification framework, illustrated in Fig.~\ref{fig:prop_flow}. 
The proposed verification flow is depicted in Fig.~\ref{fig:prop_flow}(a), while the property generation flow is shown in Fig.~\ref{fig:prop_flow}(b). The property generation flow consists of three steps: Specification Extraction, Property Definition, and Property Generation. 
In this framework, the verification process is broken down into several steps, each considering different abstraction models. A detailed introduction is provided for each step in this section.

\subsection{Specification Extraction}

\begin{figure}
	\centering
	\includegraphics{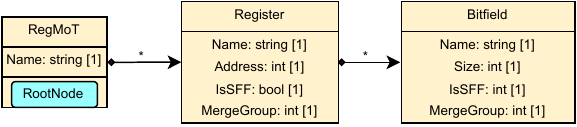}
	\caption{Simplified model for specification extraction.}\label{fig:mot}
\end{figure}

The specifications of safety registers are defined in an XML file, and all safety register information will be extracted from this file based on a pre-defined model, as shown in Fig.~\ref{fig:mot}, which is comparable to the technique utilized in our register generator~\cite{zhang2024automated}. The extracted information is stored in a specification instance, named the Model of Things (MoT), for subsequent steps. Additionally, some extra options are used for the property generator. For example, the area optimization algorithm is defined in the options. Since different optimization algorithms generate different safety register instances, it is crucial to consider the chosen algorithm during property generation. Therefore, the optimization results, which specify which safety registers are merged into which groups, is also part of the specification instances.
It is important to highlight that the same optimization algorithm must be implemented twice by both the design and verification teams, as the specification only includes the algorithm type. 
During this phase, our framework supports the checking of the optimization results between the algorithm implementations of both the design and verification teams. This capability can be used to detect algorithm bugs at a very early stage.

\subsection{Property Definition}

\subsubsection{Utilized Property Classes}
As area optimization results in completely different safety register instantiations when compared with the initial designs, the generated properties should also differ. In this section, we will discuss multiple property classes in both area-optimized and non-area-optimized modes in the Model of Properties~(MoP). The property classes utilized in this work include the following:
\begin{itemize}
\item Configurable Parameters: This property checks if the instantiation of safety components is correct, ensuring that parameters such as data width and protection method match the specifications.
\item Safety Mode: This class evaluates mainly the behavior of safety components under fault injection conditions. For safety registers, an error signal should be asserted when faults are injected. The number of faults is determined by the protection method.
\item Connectivity: This property class verifies the connectivity of safety components. For example, it focuses on the input and output connectivity of safety registers with primary inputs and outputs.
\item Normal Mode: This class is used to check the basic functionality of safety components, such as the basic register functionality.
\item Self-Test Mode: This class checks the self-test features of safety registers. During this phase, fault injection is also considered.
\end{itemize}

Due to page limitations, we cannot present all the properties used in our projects. Therefore, we will focus on three classes—Configurable Parameters, Safety Mode, and Connectivity—that are most relevant to area-optimized designs to demonstrate the property differences between original and optimized designs.

\subsubsection{Comparison between Original and Area-Optimized Designs}

\begin{figure}[h]

\begin{subfigure}{0.49\textwidth}
\begin{lstlisting}[label=lst:sva]
property param_dw_safety_registers;
    1 |-> 
    sff_wrapper_1.dw == 2 &&
    sff_wrapper_2.dw == 2 &&
    sff_wrapper_3.dw == 30;
endproperty
\end{lstlisting}
    \caption{Parameter Checking - Original Design}
    \label{fig:first}
\end{subfigure}
\hfill
\begin{subfigure}{0.49\textwidth}
\begin{lstlisting}[label=lst:sva]
property fault_injection_safety_registers;
    // Num. fault patterns decided by its data width
    // Three properties generated
    `Fault_Injection_i |-> 
    sff_wrapper_i.err_o == 1;
endproperty
\end{lstlisting}
\caption{Fault Injection Checking - Original Design}
\label{fig:second}
\end{subfigure}
\hfill
\begin{subfigure}{0.49\textwidth}
\begin{lstlisting}[label=lst:sva]
property param_dw_safety_registers;
    1 |-> 
    sff_wrapper_1.dw == 32 &&
    sff_wrapper_2.dw == 2;;
endproperty
\end{lstlisting}
    \caption{Parameter Checking - Optimized Design 1}
    \label{fig:second}
\end{subfigure}
\hfill
\begin{subfigure}{0.49\textwidth}
\begin{lstlisting}[label=lst:sva]
property fault_injection_safety_registers;
    // Two properties generated
    `Fault_Injection_i |-> 
    sff_wrapper_i.err_o == 1;
endproperty
\end{lstlisting}
    \caption{Fault Injection Checking - Optimized Design 1}
    \label{fig:first}
\end{subfigure}

\hfill
\begin{subfigure}{0.49\textwidth}
\begin{lstlisting}[label=lst:sva]
property param_dw_safety_registers;
    1 |-> 
    sff_wrapper_1.dw == 17 &&
    sff_wrapper_2.dw == 17;
endproperty
\end{lstlisting}
    \caption{Parameter Checking - Optimized Design 2}
    \label{fig:second}
\end{subfigure}
\hfill
\begin{subfigure}{0.49\textwidth}
\begin{lstlisting}[label=lst:sva]
property fault_injection_safety_registers;
    // Two properties generated
    `Fault_Injection_i |-> 
    sff_wrapper_i.err_o == 1;
endproperty

\end{lstlisting}
    \caption{Fault Injection Checking - Optimized Design 2}
    \label{fig:first}
\end{subfigure}
        
\caption{Comparison of connect properties for original and optimized designs.}\label{fig:sva_properties1}
\end{figure}

The example properties of the Parameters, Safety Mode, and Connectivity property classes are illustrated in Fig.~\ref{fig:sva_properties1} and Fig.~\ref{fig:sva_properties} for the example designs in Fig.~\ref{fig:example_opt}. In Fig.~\ref{fig:sva_properties1}, the parameter and fault-injection properties are shown. Fig.~\ref{fig:sva_properties1}(a), (c), and (e) demonstrate the properties for parameter checking. In the original design, there are three registers instantiated, so all the data width parameters must be checked. However, with area optimization techniques, the number of safety registers is reduced, resulting in different generated properties based on the utilized algorithm. Since our property generator also implements the optimization algorithm internally, these merged data width values can be determined, thus updating the properties automatically.
Additionally, in Fig.~\ref{fig:sva_properties1}(b), (d), and (f), the safety mode properties, specifically fault-injection properties, are demonstrated for both original and area-optimized designs. Similar to parameter checking, the original design requires three properties to cover the fault tolerance features due to the instantiation of three safety registers, whereas area-optimized designs require only two properties. Although the number of properties for area-optimized designs is reduced in fault-injection mode, the proving time for each property increases. This is because a merged safety register implies a larger data width, resulting in a significant increase in the number of error states considered during the verification phase.
The number of error states to be considered for EDCECC-based implementations can be formalized as follows:
\begin{equation}
    N = 2^{N_d} \cdot  \sum_{i=1}^{N_e} {N_d + N_r \choose i} ,
\end{equation}
where $N_d$, $N_r$ and $N_e$ represent the number of data bits, redundancy bits and the number of supported errors, respectively. Based on this equation, it is evident that a larger data width leads to a significant increase in the state space. If the size of the merged register is too large, the state explosion will result in an unbounded proof, which should be avoided. Therefore, in our project, we introduce a constraint on the maximum allowed merging size from the verification perspective. The verification time for both the original and optimized designs will be discussed in detail in the results section.

\begin{figure}[h]

\begin{subfigure}{0.49\textwidth}
\begin{lstlisting}[label=lst:sva]
property conn_top_input_safety_registers;
    1 |-> 
    top.a_i[1:0] == sff_wrapper_1.d_i[1:0] &&
    top.b_i[1:0] == sff_wrapper_2.d_i[1:0] &&
    top.c_i[29:0] == sff_wrapper_3.d_i[29:0];
endproperty
\end{lstlisting}
    \caption{Input Connectivity Checking - Original Design}
    \label{fig:first}
\end{subfigure}
\hfill
\begin{subfigure}{0.49\textwidth}
\begin{lstlisting}[label=lst:sva]
property conn_top_output_safety_registers;
    1 |-> 
    sff_wrapper_1.d_o[1:0] == top.a_o[1:0] &&
    sff_wrapper_2.d_o[1:0] == top.b_o[1:0] &&
    sff_wrapper_3.d_o[29:0] == top.c_o[29:0];
endproperty
\end{lstlisting}
\caption{Output Connectivity Checking - Original Design}
\label{fig:second}
\end{subfigure}
\hfill
\begin{subfigure}{0.49\textwidth}
\begin{lstlisting}[label=lst:sva]
property conn_top_input_safety_registers_optimized_1;
    1 |-> 
    top.c_i[29:0] == inst_wrapper_1.d_i[29:0] &&
    top.a_i[1:0] == inst_wrapper_1.d_i[31:30] &&
    top.b_i[1:0] == inst_wrapper_2.d_i[1:0];
endproperty
\end{lstlisting}
    \caption{Input Connectivity Checking - Optimized Design 1}
    \label{fig:second}
\end{subfigure}
\hfill
\begin{subfigure}{0.49\textwidth}
\begin{lstlisting}[label=lst:sva]
property conn_top_output_safety_registers_optimized_1;
    1 |-> 
    sff_wrapper_1.d_o[29:0] == top.c_o[29:0] &&
    sff_wrapper_1.d_o[31:30] == top.a_o[1:0] &&
    sff_wrapper_2.d_o[1:0] == top.b_o[1:0] ;
endproperty
\end{lstlisting}
    \caption{Output Connectivity Checking - Optimized Design 1}
    \label{fig:first}
\end{subfigure}

\hfill
\begin{subfigure}{0.49\textwidth}
\begin{lstlisting}[label=lst:sva]
property conn_top_input_safety_registers_optimized_2;
    1 |-> 
    top.c_i[16:0] == inst_wrapper_1.d_i[16:0] &&
    top.c_i[29:17] == inst_wrapper_2.d_i[12:0] &&
    top.a_i[1:0] == inst_wrapper_2.d_i[14:13] &&
    top.b_i[1:0] == inst_wrapper_2.d_i[16:15];
endproperty
\end{lstlisting}
    \caption{Input Connectivity Checking - Optimized Design 2}
    \label{fig:second}
\end{subfigure}
\hfill
\begin{subfigure}{0.49\textwidth}
\begin{lstlisting}[label=lst:sva]
property conn_top_output_safety_registers_optimized_2;
    1 |-> 
    sff_wrapper_1.d_o[16:0] == top.c_o[16:0] &&
    sff_wrapper_2.d_o[12:0] == top.c_o[29:17] &&
    sff_wrapper_2.d_o[14:13] == top.a_o[1:0] &&
    sff_wrapper_2.d_o[16:15] == top.b_o[1:0];
endproperty

\end{lstlisting}
    \caption{Output Connectivity Checking - Optimized Design 2}
    \label{fig:first}
\end{subfigure}
        
\caption{Comparison of connect properties for original and optimized designs.}\label{fig:sva_properties}
\end{figure}

Apart from the parameter and fault injection checking, the connectivity check is also critical for the optimized designs. The input and output connectivity checking for the original design is presented in Fig.~\ref{fig:sva_properties}(a) and (b).
However, when the smaller registers are merged, the one-to-one mapping relationship between the safety registers and the original registers disappears. Consequently, the connectivity checking properties must be updated based on the optimization algorithms, making verification challenging even for small designs. Our proposed verification framework can automatically generate the adapted properties without requiring any manual effort. The adapted properties for the optimized designs can be found in Fig.~\ref{fig:sva_properties}(c) and (d) for input connectivity checking and in Fig.~\ref{fig:sva_properties}(d) and (f) for output connectivity checking. 


\subsection{Property Generation}\label{sec:framework:mov}
After defining all property classes, the properties can be generated based on different view requirements in the Model of View~(MoV). In this work, we mainly focus on the use of System Verilog Assertions (SVAs). With these assertion files and the design files, the formal tool can be triggered to obtain the verification results.


\section{Results}\label{sec:results}

\subsection{Setup}
In this work, we applied our proposed framework using safety register data from three automotive SoCs. This property generation framework is based on Python, with the area optimization algorithms embedded in the same generator. The ILP solver used in this work is PuLP~\cite{pulp}. The generated properties are verified using a commercial property checking tool.

\subsection{Formal Verification Results}
\subsubsection{Verification Time}

\begin{table}
\centering
\begin{tabular}
{|c|c|cc|}
\hline
&  \multicolumn{1}{c|}{Initial Design} & \multicolumn{2}{c|}{Verification Time Overhead} \\

     & \#SafetyReg    & \multicolumn{1}{c}{BFD (32-bit)} & \multicolumn{1}{c|}{ILP (32-bit)}  \\
\hline
SoC1 &  1827 &  6.53\%     & 14.87\%                   \\
\hline
SoC2  & 2392  & 71.05\%        & 84.32\%                      \\
\hline
SoC3  & 49  &  144.27\%        & 131.02\%                      \\
\hline
\end{tabular}
\caption{Comparison of verification time.}\label{tb:ver_time}
\vspace{-20pt}
\end{table}


Although the same property classes are defined for both the original and optimized designs, the generated properties differ based on the chosen algorithm, which results in varying verification times across different designs. In particular, the safety mode properties can consume significantly more time in the optimized designs compared to the original designs due to the larger data width of the merged registers. In Table~\ref{tb:ver_time}, we present the overhead in verification time (property proving time) for three SoCs with the maximum allowed register size set to 32 during the optimization. In this work, all properties can still converge because of the constraints provided by the verification team. From this table, it is evident that the verification time overhead varies across different SoCs and different algorithms. Based on our observations, if small-sized safety registers, such as 4-bit or 8-bit, comprise a large portion of the design, the verification time overhead is significant. This is because these registers will be merged into a 32-bit safety register, and the safety-related properties consume significantly more time, as seen in the case of SoC3.

\subsubsection{Detected Bugs}

This proposed framework has been applied to a production project. Using it, we identified five bugs related to safety integration—three in connectivity and two in the area-optimized implementation—in a short period. The bugs found in the area-optimized designs were also related to connectivity.  
One bug we identified was that some bits of merged registers were not connected under specific configurations. Another bug was related to the clock and reset connections in the optimized designs.
This is because optimized designs obscure the one-to-one mapping between register specifications in the XML file and the registers instantiated in the RTL code.
Additionally, since all properties are generated automatically, the verification effort for each safety structure in this project has been reduced from five person-days (PD) to one PD.

\subsection{Comparison with Another Verification Framework}
After discussing the verification results with our proposed framework, we would like to compare it with previous work that also focuses on safety register verification. The comparison items are listed in Table~\ref{tb:comparsion}.
In the previous work, area-optimized designs are  not supported, and it still relies on a one-to-one mapping between specification and RTL code to conduct the verification. Additionally, the previous work extracts design information from RTL to construct multiple tables, allowing connectivity checking to be shifted to table checking without triggering any formal tools. Although this approach is faster, it makes debugging difficult if the check fails and a Counter-Example~(CEX) trace is required.
Moreover, the previous work covers safety register verification only partly, as it splits the verification into two phases: functional and integration. In the functional phase, a simplified safety architecture is verified for its main safety features, such as fault injection features, by an individual team. Then, integration verification is conducted based on the generated tables. In contrast, our work verifies all features with the complete register blocks. Both functional and integration aspects are verified simultaneously, ensuring comprehensive coverage and reducing the risk of overlooking any potential issues.

\begin{table}
\centering
\begin{tabular}
{|c|c|c|c|c|}
\hline
         & Area-Opt. & CEX Trace & Functional & Integration \\
\hline
This Work & yes       & yes          & yes            & yes             \\
\hline
Work~\cite{busch2023integration}  & no        & no          & no             & yes             \\
\hline
\end{tabular}
\caption{Comparison between This Work and Previous Work~\cite{busch2023integration}.}\label{tb:comparsion}
\vspace{-20pt}
\end{table}

\section{conclusion}\label{sec:conclusion}

Safety registers are critical components in modern automotive SoCs, requiring extensive verification to ensure correct behavior. Safety features often introduce significant area overhead, making area-optimized approaches essential for more efficient automotive SoCs. However, optimizing for area introduces additional verification challenges, as optimized designs may not naturally adhere to the initial specifications.
To address these verification challenges, we propose an automated formal verification framework. This framework automatically generates properties based on register specifications and the chosen optimization algorithm. These properties can be seamlessly used with commercial formal verification tools. The framework has been applied to a production project, significantly reducing human effort and enabling the discovery of bugs in area-optimized designs within a short time period. 
Lastly, we would like to highlight that this verification approach is not limited to our in-house SFF library. The verification flow can be adapted for other safety libraries as well, providing a versatile and scalable solution for safety register verification across different projects and platforms.

\section*{Acknowledgment}
We also want to thank our colleagues, Basavaraj Naik, Vidya Sagar Kantamneni, Yifang Wang, Xin An and Djones Lettnin, for their valuable input and suggestions.

\bibliographystyle{IEEEtran}
\bibliography{bibfile}
\end{document}